\documentclass[12 pt, a4paper]{article}
\usepackage{physics}
\usepackage{jheppub}
\usepackage{bm}
\usepackage{xcolor}
\usepackage{amsmath}
\usepackage{amsfonts}
\usepackage{url}
\usepackage{arydshln}
\usepackage{chngcntr}
\counterwithout{equation}{section}

\usepackage{lmodern}
\usepackage[utf8]{inputenc}
\usepackage[T1]{fontenc}
\definecolor{refkey}{gray}{0.45}
\definecolor{labelkey}{RGB}{155,48,48}
\hypersetup{
  colorlinks=true,
  citecolor=magenta,
  linkcolor=blue,
  urlcolor=violet,
  pdftitle={Upamanyu Moitra: Heterotic Black Holes in Duality-Invariant Formalism},
  pdfauthor={Upamanyu Moitra},
 }
 
 \title{\center{ \fontfamily{lmr}\selectfont Heterotic Black Holes in Duality-Invariant Formalism}}

\author{\center \fontfamily{lmr}\selectfont Upamanyu Moitra}

\affiliation{ 
\begin{center}
Institute for Theoretical Physics,  Institute of Physics, Universiteit van Amsterdam, Science Park 904, 1089 XH
Amsterdam, The Netherlands

\href{mailto:u.moitra@uva.nl}{\textup{\texttt{u.moitra@uva.nl}}}
\end{center}
}

\abstract{ {\centering  We consider the effective theory of heterotic strings in two spacetime dimensions,  in a double field theory-inspired formalism,  manifestly consistent with $T$-duality in string theory.   Restricting the gauge group to a single $\mathrm{U}(1)$,  we study the charged black hole solution and perform a precise analysis of the properties of the dual geometry with the $\mathrm{O}(1,2; \mathbb{R})$-valued generalized metric.  We comment on some aspects related to singularities and gauge dependence.  We show that the classification program for higher-derivative corrections can also be applied to the heterotic case.  We further elucidate how a previously proposed solution to the equations of motion,  parametrized in a manner fully non-perturbative in $\alpha'$,  can be extended to the scenario with $r$ abelian fields and the corresponding $\mathrm{O}(1,1+r; \mathbb{R})$ symmetry.  We discuss some novel features of the solution for charged black holes.  } }

\newcommand{\ldc}{\xi}
\newcommand{\eig}{\rho}
\newcommand{\cS}{\mathcal{S}}
\newcommand{\cD}{\mathcal{D}}

\newcommand{\mi}{\mathrm{i}}

\newcommand{\N}{\nabla}

\newcommand{\bee}{\begin{equation}}
\newcommand{\ee}{\end{equation}}
\newcommand{\beq}{\begin{equation*}}
\newcommand{\eeq}{\end{equation*}}
\newcommand{\baa}{\begin{equation}\begin{aligned}}
\newcommand{\ea}{\end{aligned}\end{equation}}

\makeatletter
\gdef\@fpheader{}
\makeatother

\begin{document}
\maketitle

\section{Introduction}\label{sec-intro}

The philosophy of reductionism in physics aims to describe nature with a minimal set of fundamental principles. 
The standard model of elementary particles, for instance, has been extraordinarily successful in describing all observed non-gravitational phenomena. 
It has, however, been appreciated for a long time that there is nothing elementary about elementary particles. 
The phenomenon of duality --- which can be taken as a strengthening of reductionism --- in quantum field theory demonstrates that what appears fundamental in a certain duality frame might not be fundamental in another.  
Various dualities play a central role in our current understanding of string theory. 
One particularly remarkable example is that of target space duality or $T$-duality \cite{Giveon:1994fu}, which makes the astonishing assertion that physics, as probed by strings, is completely equivalent for a circle of size $R$ and another circle of size $\alpha'/R$,  where $\sqrt{\alpha'}$ is the string length.

A consideration of $T$-duality in non-trivial backgrounds provides an important window into how strings perceive spacetime itself. 
The transformation laws of background fields corresponding to massless string states are specified by the Buscher rules \cite{Buscher:1987sk, Buscher:1987qj} and their generalization. 
Studying the consequences of $T$-duality in cosmological backgrounds has led to a number of surprising insights \cite{Brandenberger:1988aj, Tseytlin:1991xk, Veneziano:1991ek}. 
As a particularly striking consequence of $T$-duality,  the low-energy effective action of string theory, compactified on a $d$-dimensional torus, enjoys an $\mathrm{O}(d, d; \mathbb{R})$ symmetry \cite{Meissner:1991zj, Sen:1991zi, Gasperini:1991ak, Maharana:1992my}. 
While the continuous symmetry is broken to its discrete subgroup $\mathrm{O}(d, d; \mathbb{Z})$ in the full quantum theory, the $\mathrm{O}(d, d; \mathbb{R})$ group is known to be an exact symmetry of classical string theory, to \emph{all} orders in $\alpha'$. 
Nevertheless, writing down a low-energy effective action in an $\alpha'$ expansion consistent with this symmetry is an extremely non-trivial task. 
It is well known that generic effective field theories fail to capture some characteristic features of string theory.
The elegant formalism of double field theory (DFT) \cite{Siegel:1993th, Hull:2009mi} has proven to be very useful in this regard. 
DFT makes $T$-duality, a very stringy phenomenon, manifest in the field theory language, at the cost of doubling the number of compact coordinates, see \cite{Aldazabal:2013sca} for a review.

In the recent years, DFT-inspired methods have led to a progress in understanding the nature of higher-derivative corrections in several contexts,  including cosmological backgrounds with all the spatial dimensions compactified. 
The authors of \cite{Hohm:2019jgu} were able to provide a complete classification of all the higher-derivative terms in the full effective action, up to field redefinitions. 
More precisely, they found a field basis in which the most general action assumes a particularly simple form, correct to an arbitrary order in the derivative or $\alpha'$ expansion (these two expansions are not always equivalent, as in the case explored in this paper;  see \cite{Codina:2023fhy} for more comments). 
Provided the existence of such a complete classification, one can seriously entertain the possibility that the resulting action gives a non-perturbative (in $\alpha'$) description of string backgrounds.

A closely related scenario of the cosmological set-up is that of non-critical strings in two spacetime dimensions, for which there exist interesting black hole solutions \cite{Elitzur:1990ubs, Rocek:1991vk, Mandal:1991tz, Witten:1991yr}. 
In this case, one performs the dualization procedure along the temporal direction. 
Some aspects of this duality were pointed out, for instance, in \cite{Giveon:1991sy, Tseytlin:1991wr, Dijkgraaf:1991ba, Kiritsis:1991zt}. 
The cosmological classification program of \cite{Hohm:2019jgu} was extended to 2D black holes in \cite{Codina:2023fhy}, where the authors presented the general action in terms of the dilaton and the metric (the Kalb-Ramond 2-form is trivial in two dimensions). 
However, a very interesting and physically rich possibility seems to have been neglected so far in this context: there can be gauge fields in the massless sector of 2D string theory. 
The purpose of this note is to fill this gap in the literature.

It is known \cite{McGuigan:1991qp} that two consistent heterotic string theories exist in two spacetime dimensions, with gauge fields in the massless sector, the possible gauge groups being $E_8 \times \mathrm{Spin}(8)$ and $\mathrm{Spin}(24)$. 
If one considers a single abelian subsector of the gauge groups, the low-energy effective description includes an additional Maxwell field. 
The authors of \cite{McGuigan:1991qp} had also discussed the charged generalization of the two-dimensional black hole. 
How can we extend the discussions of duality and higher-derivative corrections in this scenario? 
For the theory with $r$ abelian gauge fields compactified on a $d$-torus, the symmetry group is $\mathrm{O}(d, d+r; \mathbb{R})$. 
There exists a formulation of DFT in the context of heterotic strings \cite{Hohm:2011ex}, which we can use to describe the two-dimensional charged black hole in a duality-invariant manner.

In \S\ref{sec-twoder}, we rewrite the two-dimensional effective action up to two derivatives for time-independent backgrounds in a manifestly duality-invariant manner and discuss the charged black hole solution first discovered in \cite{McGuigan:1991qp}. 
Then we focus our attention on the duality transformations of the solution and write down the precise analogs of the Buscher rules in this case. 
The metric and gauge field components are mixed up in the duality transformations in a very interesting manner, which leads to some intriguing aspects of the black hole geometry and its dual, not previously discussed in the literature. 
We also point out some issues that arise vis-\`{a}-vis gauge invariance and duality.

In \S\ref{sec-solun} we undertake the task of providing a higher-derivative generalization of the action, as in \cite{Hohm:2019jgu, Codina:2023fhy}. 
We find that this procedure can be conducted in a  similar straightforward manner and clarify some aspects of the generalization in this process. 
The gauge field and the metric appear in a coordinated fashion in the full action,  with only one independent Wilson coefficient appearing at each order in derivative.
We next turn our attention to the question of solutions to the equations of motion in the full $\alpha'$-corrected theory in a non-perturbative manner. In \cite{Codina:2023nwz}, the authors discussed a very useful parametrization, first found by  \cite{Gasperini:2023tus}, to describe such solutions. 
We find, rather remarkably, that this parametrization can also be extended to our scenario involving the gauge field. 
We discuss how the charged black hole in the full theory can be incorporated in this framework and some novel features associated with the solution,  including an issue that arises in the extremal limit.  

We discuss in \S\ref{sec-discussion} the underlying reason as to why the aforementioned parametrization is possible even with an additional gauge field and show that the result can be extended to the situation with $r$ abelian gauge fields with the corresponding $\mathrm{O}(1,1+r; \mathbb{R})$ symmetry.  We conclude this note with a discussion on some future prospects.

\section{Duality in Two-Derivative Heterotic Black Holes}\label{sec-twoder}
\subsection{Two-Derivative Black Hole Solution}\label{subsec-2dbh}

We first look at the two-derivative heterotic theory with a single $\mathrm{U}(1)$ gauge field turned on and explore some interesting consequences of duality even at this level.  The two-derivative action describing the ``massless'' sector is given by
\baa
S^{(2)} = \frac{1}{16\pi G} \int \dd[2]{x} \sqrt{-g} \, e^{-2\phi} \pqty{ \ldc^2 + R + 4(\N \phi)^2 - \frac{1}{4} F_{\mu \nu} F^{\mu \nu}  }. \label{act1}
\ea
In the critical dimension,  one has $\ldc = 0$; in the present discussion,  we require $\ldc^2$ to be positive, but its precise value would not play an important role.   In the  context of 2D heterotic strings \cite{McGuigan:1991qp},  we have $\ldc^2 = 8 /\alpha'$.  As in \cite{Codina:2023fhy,  Codina:2023nwz} we consider time-independent solutions of the  form
\baa
\dd{s}^2 &= -m(x)^2 \dd{t}^2 + n(x)^2 \dd{x}^2, \\
\phi &= \phi(x),  \\
A_\mu \dd{x^\mu} &= V(x) \dd{t},  \label{ansatz}
\ea
where we have introduced an appropriate time-independent ansatz for the gauge field\footnote{Note that a possible term of the form $P(x) \dd{x}$ in the gauge field $A$ can be eliminated by a gauge transformation.}.  Using this reduction ansatz and ``integrating over'' the temporal coordinate,  we find the following one-dimensional action,
\baa
S^{(2)} = \frac{1}{16\pi G} \int \dd{x} m n e^{-\phi(x) }  \bqty{ \ldc^2 - \frac{2}{m(x)n(x)} \dv{x} \frac{m'(x)}{n(x)} + \frac{4\phi'(x)^2  }{n(x)^2} + \frac{V'(x)^2}{2m(x)^2 n(x)^2}  }, \label{action2}
\ea
where we have absorbed the volume of the temporal coordinate in $G$.

It might be hard to see from the form \eqref{action2} that the Lagrangian enjoys an $\mathrm{O} (1, 2; \mathbb{R})$ symmetry.  We will use insights from double field theory to cast this action in a manifestly symmetric form.  For this purpose,  as is usual,  we define the duality-invariant dilaton field,
\baa
\Phi (x) = 2\phi(x) - \log | m(x) |.  \label{dindil}
\ea
The generalized metric describing the metric and gauge field \cite{Hohm:2011ex}  in this case is a $3 \times 3$ symmetric matrix
\baa
{\cal H} = \begin{pmatrix}
- \dfrac{1}{m(x)^{2}} &&&  \dfrac{V(x)^2}{2m(x)^2}  &&& \dfrac{V(x)}{m(x)^2}  \\
 \dfrac{V(x)^2}{2m(x)^2}  &&&  - \dfrac{\pqty{2m(x)^2 - V(x)^2}^2}{4m(x)^2} &&& V(x) - \dfrac{ V(x)^3}{2m(x)^2} \\
 \dfrac{V(x)}{m(x)^2}  &&& V(x) - \dfrac{ V(x)^3}{2m(x)^2} &&& 1 -  \dfrac{V(x)^2}{m(x)^2} 
\end{pmatrix}. \label{genmet1}
\ea
It is easy to see that ${\cal H}$ is an element of $\mathrm{O} (1, 2; \mathbb{R})$: it preserves the quadratic form $\eta$ of signature $(1,2)$,
\baa
\eta = \begin{pmatrix}
0 & 1 & 0 \\
1 & 0 & 0 \\
0 & 0 & 1
\end{pmatrix},  \qquad {\cal H} \eta {\cal H}^{\rm T}  = \eta.  \label{defqform}
\ea
In previous works \cite{Codina:2023fhy,  Codina:2023nwz} where the gauge field was absent $V(x) = 0$,  only the first $2 \times 2$ diagonal blocks of the matrices ${\cal H}$ and $\eta$ were pertinent to the discussion.  As we shall soon see,  the interplay between the gauge field and the spacetime metric  gives rise to rather interesting physics.  As in these works, it is actually more convenient to work with the field variable
\baa
\cS \equiv \eta \mathcal{H}.  \label{defcalS}
\ea
The matrix $\cal S$ is also valued in $\mathrm{O} (1, 2; \mathbb{R})$ but is not symmetric.
It is worth noting that the function $n(x)$ in \eqref{ansatz} changes under a reparametrization of the spatial coordinate $x$.  This motivates one to define a reparametrization-covariant derivative,
\baa
\cD \equiv \frac{1}{n(x)} \dv{x}, \label{defcovD}
\ea
whose properties can be found in \cite{Hohm:2019jgu,  Codina:2023fhy}.

Using the duality-invariant dilaton \eqref{dindil},  the action \eqref{action2} can be put in the form
\baa
S^{(2)} &= \frac{1}{16\pi G} \int \dd{x} \frac{e^{ - \Phi }}{n} \pqty{ \ldc^2 n^2 + \Phi'^2  - \frac{2m'^2 - V'^2}{2m^2} } \\
&= \frac{1}{16\pi G} \int \dd{x} n e^{-\Phi}  \pqty{ \ldc^2 + (\cD \Phi)^2  + \frac{1}{8} \Tr (\cD \cS)^2  }.  \label{action12}
\ea
Note that we have performed an integration by parts to obtain the form in the first line. In the second line, we have cast the action in a manifestly $\mathrm{O} (1, 2; \mathbb{R})$-invariant and reparametrization-invariant form.  Under the action of $\mathrm{O} (1, 2; \mathbb{R})$,  the fields $n$ and $\Phi$ are singlets and $\cS$ is a rank-2 tensor.  The compact form of the action presented in the second line is the same whether or not the gauge field is present,  which is a feature of the elegant formalism.

Let us obtain the charged black hole solution from the equations of motion by varying the fields $n$,  $\Phi$,  $m$ and $V$.  
It is actually instructive to go through the steps in some detail. 
Denoting $E_X \equiv 16\pi G \fdv*{S}{X}$ for the field $X$, we find the equations of motion,
\baa
E_n &= - \frac{ e^{-\Phi} }{n^2} \pqty{ \Phi'^2 - \frac{2m'^2 - V'^2}{2m^2} - \ldc^2  n^2 } = 0 , \\
E_\Phi &= -  \frac{ e^{-\Phi} }{n} \pqty{ \Phi'^2 - \frac{2m'^2 - V'^2}{2m^2} + \ldc^2 n^2   } - 2 \dv{x} \pqty{ \frac{ e^{-\Phi} }{n}  \Phi' } = 0 ,  \\
E_m &= \frac{e^{-\Phi} }{n}  \frac{2m'^2 - V'^2}{m^3} + 2\dv{x} \pqty{  \frac{e^{-\Phi} }{n}  \frac{m'}{m^2} } = 0, \\
E_V &= - \dv{x} \pqty{ \frac{e^{-\Phi} }{n}  \frac{V'}{m^2}  } = 0.  \label{eoms1}
\ea
In the Schwarzschild-Droste--like gauge $n(x) = 1/m(x)$, the combination of the equations $n(x) E_n + E_\Phi - m(x) E_m = 0$ yields
\baa
2m'^2 - 2m(m''+ m \Phi'') = 0,  \label{simpphieom}
\ea
which can be solved immediately to give
\baa
\Phi(x) = \Phi_0 - b x - \log | m(x) | ,  \label{phisol}
\ea
where $\Phi_0$ and $b$ are integration constants. Inserting this in the $E_V  = 0$ equation,  we find that
\baa
V(x) = -  \sqrt{2} Q  e^{- b x},  \label{gaugesol1}
\ea
where $Q$ is a dimensionless integration constant,  related to the total charge in the system. 
The $E_n = 0$ equation can now be solved to give
\baa
m(x) =\sqrt{ \frac{\ldc^2}{b^2} - 2 \mu e^{-b x} + Q^2 e^{-2 b x}   }.  \label{mxsol2}
\ea
Here, $\mu>0$ is another dimensionless integration constant, related to the mass of the black hole.  The geometry is of course asymptotically flat as $x \to \infty$ --- the requirement that asymptotically $x$  has the same scale as the standard flat coordinate fixes $b = \ldc$. This is exactly the solution obtained by the authors of \cite{McGuigan:1991qp}. The physical mass and charge are given by $(16\pi G)^{-1} 2 \ldc \mu e^{-\Phi_0}$ and $(16\pi G)^{-1} \sqrt{2} \ldc Q e^{-\Phi_0}$ respectively.  We can set $\Phi _0 = 0$ with impunity since we can absorb any translation of the dilaton field in the two-dimensional Newton constant $G$.   Note that out of the four equations in \eqref{eoms1}, we have used only three (combinations) to arrive at the solution --- the one remaining independent combination of the equations is automatically satisfied by this solution.  This happens because the four equations are not independent and related by a Bianchi identity \cite{Moitra:2022umq, Codina:2023fhy}.   

This geometry has two horizons in general, which are located at $x = x_\pm$, for which $m(x)^2 = 0$,
\baa
x_\pm = \frac{1}{\ldc} \log( \mu \pm \sqrt{\mu^2 - Q^2} ),  \label{xhorilox}
\ea
with the upper (lower) sign corresponding to the event (Cauchy) horizon for $0\leq |Q| \leq \mu$,  the parameter range we consider throughout.  The function $m(x)^2$ is negative for $x_- < x< x_+$.  While the duality-invariant dilaton $\Phi$ blows up near the horizons,   the string dilaton $\phi$ is perfectly well-behaved.  The scalar curvature $R$,  the norm of the electric field strength $F_{\mu \nu} F^{\mu \nu}$ and the string coupling $g_s = e^{\phi}$ blow up as $x \to -\infty$, however as explained in \cite{Moitra:2022umq},  this is not a true singularity in the sense that it cannot be reached by a null geodesic in a finite affine time.  In the extremal limit $|Q| \to \mu$,  the two horizons coalesce into one as a degenerate horizon.

\subsection[The $T$-Dual Geometry]{\boldmath The $T$-Dual Geometry}
\label{subsec-tdual}

We now focus on the properties of another geometry which is related to the black hole geometry by a duality transformation.   We are interested in particular in the $T$-dual of this geometry and in stating the Buscher rules  which characterize it.  The generalized metric formulation of double field theory is very powerful in this regard. Given the ${\cal H}$ defined in \eqref{genmet1},  we can obtain the generalized metric $\widetilde{\cal H}$ for the $T$-dual  configuration by simply conjugating it with $\eta$, 
\baa
\widetilde{\cal H} = \eta {\cal H} \eta. \label{htilde}
\ea
Denoting the fields in the dual geometry by a tilde,  we find that the metric component and the gauge field are related as follows,
\baa
\widetilde{m}(x)^2 &= \frac{4m(x)^2}{(2m(x)^2 - V(x)^2)^2}, \\
\widetilde{V}(x) &= \frac{2V(x)}{2m(x)^2 - V(x)^2}. \label{dualmvx}
\ea
On the other hand the fields $n$ and $\Phi$ are duality-invariant,
\baa
\widetilde{n}(x) = n(x), \qquad \widetilde{\Phi}(x) = \Phi (x).  \label{nphiunch}
\ea
The quantity $V(x)^2 / m(x)^2$ is also a duality-invariant.  The duality transformation \eqref{dualmvx} which mixes the metric and gauge field in a rather non-trivial manner could be rather surprising at first sight,  but the generalized metric formalism makes this obvious. From the point of view of the equations of motion \eqref{eoms1},  the $E_n$ and $E_\Phi$ equations remain invariant,  but the remaining two equations can be written in the form
\baa
E_{\widetilde{m}} &= -\frac{1}{2} \pqty{ 2m(x)^2 + V(x)^2 } E_m - 2m(x) V(x) E_V \\
E_{\widetilde{V}} &=  m(x) V(x) E_m + \frac{1}{2} \pqty{ 2m(x)^2 + V(x)^2 } E_V, \label{rotatedeom}
\ea
which makes it clear that if $m$, $n$,  $\Phi$ and $V$ solve the equations of motion so do $\widetilde{m}$,  $\widetilde{n}$ $\widetilde{\Phi}$ and $\widetilde{V}$. Let us write down the full solution obtained in the previous subsection,
\baa
m(x)^2 &= 1 - 2\mu e^{-\ldc x} + Q^2 e^{-2\ldc x} = n(x)^{-2},  \\
\Phi(x) &= - \ldc x -\frac12 \log| 1 - 2\mu e^{-\ldc x} + Q^2 e^{-2\ldc x} |, \\
V(x) &=   -  \sqrt{2} Q  e^{- \ldc x}.  \label{origsols}
\ea
We immediately find from \eqref{dualmvx},
\baa
\widetilde{m} (x)^2 &= \frac{1 - 2\mu e^{-\ldc x} + Q^2 e^{-2\ldc x} }{ \pqty{1 - 2\mu e^{-\ldc x} }^2 } , \\
\widetilde{V} (x) &= -\frac{ \sqrt{2} Q  e^{- \ldc x}}{1 - 2\mu e^{-\ldc x}}. \label{dualmvxs1}
\ea

The point of view we espouse in this paper corresponds to that of the standard discussion of $T$-duality --- we consider the fields in the two different duality frames to describe different spacetimes and gauge field configurations.  They are as different as two circles with radii $R$ and $\alpha'/R$.  Duality then corresponds to the surprising statement that physics --- as probed by strings --- in these two spaces are equivalent.  It is worth mentioning that in the context of 2D black holes,  there is another point of view of duality which relates the physics in different regions of the \emph{same} spacetime.  In this perspective, one is forced to consider spacetime regions ``behind'' the singularity.  In the context of the uncharged two-dimensional black hole,  the importance of this ``beyond-the-singularity'' region was discussed in in \cite{Witten:1991yr} and its relation to $T$-duality was  pointed out in \cite{Giveon:1991sy, Dijkgraaf:1991ba}. Some aspects of duality in charged black holes were discussed in, for example, \cite{Giveon:1991jj,  Giveon:2003ge}. The great advantage of the formalism of generalized metric is that it allows us to make the duality map very precise.

Coming back to the actual system,  note that we have chosen the gauge in such a way (more about which in the next subsection) so that in the dual geometry $\widetilde{m} (x) \to 1$ as $x \to \infty$.  However,  we notice that both $\widetilde{m} (x)$ and $\widetilde{V} (x)$ blow up at $x = x_{S} \equiv \ldc^{-1} \log(2\mu)$.  In fact,  the Ricci scalar,  electric field strength and the string coupling all blow up in the vicinity of $x = x_S$ in the dual geometry,
\baa
\widetilde{R} \sim \frac{1}{(x- x_S)^2}, \quad \widetilde{F^{\mu \nu} F_{\mu \nu} } \sim \frac{1}{(x-x_S)^2}, \quad \widetilde{e^{2 \phi}} \sim \frac{1}{(x-x_S)}.   \label{dualinv}
\ea
The presence of a singularity in the dual geometry is hardly surprising.   Such singularities have been noted in the literature for uncharged black holes.  In fact, the authors of \cite{Codina:2023nwz} had argued for the presence of a singularity whenever there is a horizon.  But in all such cases,  it is the \emph{horizon} itself that gets mapped to a curvature singularity in the dual geometry.  What is rather surprising in our case is that the singularity in the dual geometry is at a point which is \emph{outside} the black hole horizon in the original geometry $x_S > x_+$.  It is as though a string can ``feel'' a charged horizon even    from far. 

Note quite importantly that the singularity in the dual geometry is a \emph{timelike} singularity not hidden behind an event horizon (i.e.,  a naked singularity).  But is this singularity \emph{truly} a singularity? At first sight,  it would appear to be so because it appears at a finite value of the $x$ coordinate.  However,  we can easily solve the null geodesic equation to express the affine parameter for an ingoing light ray travelling between two points $x_1$ and $x_2$
\baa
\lambda (x_2) - \lambda (x_1)  = \frac{1}{\cal N } \int_{x_1}^{x_2} \dd{x}  \frac{ \widetilde{m} (x) }{ m (x) }, \label{affinepar}
\ea
where $\cal N$ is a normalization constant.
We can choose an arbitrary point $x_1 > x_S$ such that $\lambda (x_1)  = 0$.  Using the forms of the functions   in \eqref{origsols} and \eqref{dualmvx},  we find that as $x_2 \to x_S$,  $\lambda(x_2) \to \infty$ logarithmically.  Thus,  even this naked singularity is not a true singularity in this sense because it is infinitely far in affine time.  In spite of this,  we can characterize the singularity in the conventional manner by imposing a physical cut-off in the curvature or the string coupling.

\begin{figure}
\centering
\includegraphics[height=7cm]{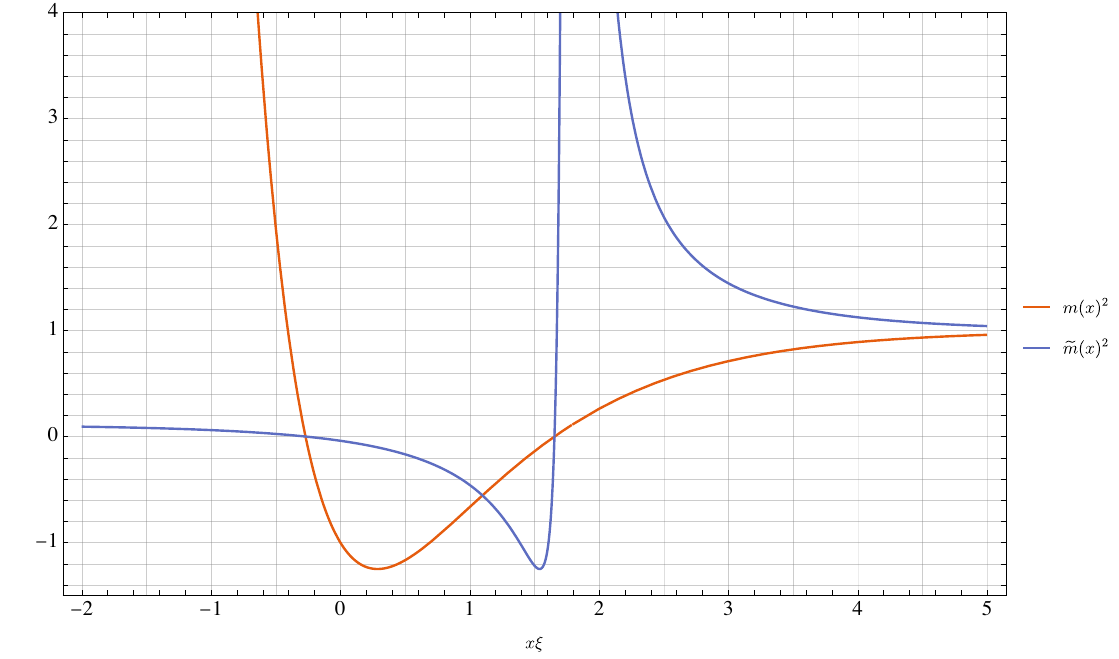}
\caption{The behavior of the metric component $-g_{tt}$ in the original and dual geometries. In both cases, $n(x) = m(x)^{-1}$.  The mass and charge parameters are chosen to be $\mu =2$ and $Q = 3$.}
\label{fig:dualms}
\end{figure}

From the large $x$ fall-off of the metric and gauge field, we see that the physical mass and charge are proportional to $-\mu$ and $Q$ respectively --- although duality preserves the value of the  charge,  it turns a spacetime with positive mass  into one with a negative mass.

If we adopt the point of view that the geometry is ``built inwards'' from an asymptotic region of small string coupling,  then the dual geometry described by \eqref{dualmvx} terminates at the singularity,. We can,  however, still wonder about the regions dual to the black hole interior.  We see from \eqref{dualmvx} that $ \widetilde{m}(x)^2 $ is negative precisely in the region where  $m(x)^2$ is negative and has the same zeros. Thus, the region between the Cauchy and event horizons is self-dual.  On the other hand,  the region between the Cauchy horizon and the timelike singularity in the original geometry gets mapped to a region which is like the exterior of the black hole in an asymptotically flat space.  The nature of these geometries can be understood more clearly from the plots of the functions $m(x)^2$ and $ \widetilde{m}(x)^2$ in the original and dual geometries, see Figure \ref{fig:dualms}.

\subsection{Gauge Dependence and Duality}\label{subsec-gauge}

The careful reader would perhaps have noticed that in writing down the solution \eqref{gaugesol1} to the equation of motion involving the gauge field, we have already made a gauge choice,  albeit a completely natural one: the gauge potential vanishes at the asymptotic infinity $x = \infty$.  One could also make a choice in which the potential attains a constant value $V_0$.
\baa
V(x) = V_0 - \sqrt{2} Q e^{ - \ldc x }. \label{nonzerov}
\ea
It is worth emphasizing that the duality transformation \eqref{dualmvx} involves the gauge field itself and \emph{not} the gauge-invariant field strength.  This naturally raises the question as to what aspects of the duality presented in the previous subsection are dependent on the choice of gauge.

The natural choice $V_0 = 0$ actually fits in nicely with the duality transformation \eqref{dualmvx} presented above: in the asymptotic region this ensures that $\widetilde{m} \to 1$ as $m \to 1$.  For a non-zero value of $V_0$,  $\widetilde{m}$ will tend to a different value asymptotically,  In this case,  the zero of the denominator of $\widetilde{m}(x)^2$  in \eqref{dualmvx} is at the value $x = x_S$,  where
\baa
x_S = \frac{1}{\ldc} \log \frac{2(2\mu - \sqrt{2} Q V_0 ) }{2 - V_0^2}.  \label{xsdefs}
\ea
It would seem that by tuning $V_0$,  we would be able to put this singularity practically at any value of $x$: outside the horizon event horizon, between the two horizons, behind the Cauchy horizon; we could even make it disappear by ensuring that $x_S$ is not real.  This is physically not a very satisfactory picture --- whether or not there is a horizon in the geometry should not depend on the gauge choice.  In this respect,  it would appear that it is sensible to impose the original gauge choice \emph{by fiat} in the sense that we should impose the same boundary conditions for the original geometry and the dual geometry.

In this context,  let us also comment briefly on another widely used gauge which is natural in Euclidean quantum gravity: we choose the value $V_0$ such that the gauge potential vanishes on the event horizon at $x = x_+$ and hence non-singular,
\baa
V_0  = \sqrt{2} Q e^{ - \ldc x_+ }.  \label{gauge2}
\ea
Quite interestingly,  in this gauge we are able to make the same sort of statement made in the previous subsection for the dual exterior region: the geometry has a  timelike ``singularity'' at $x =  x_+$ which is infinite affine distance from any other exterior point.  In this case,  the dual geometry terminates at exactly the location of the original horizon.  However, this fails in the extremal limit since in this gauge $2m(x)^2 = V(x)^2$, which implies that the denominator of $\widetilde{m} (x)^2$ vanishes \emph{everywhere}. This is another indication that  in this case we should look at only the $V_0 = 0 $ gauge.

We might be interested again in the dual of the black hole interior in the non-extremal case.  In contrast with the exterior geometry,  the result is physically quite different in this case.   The interior was, in a certain sense,  self-dual in the $V_0 = 0$ gauge.  However,  in the gauge \eqref{gauge2}, the event horizon is  mapped to a \emph{spacelike} curvature singularity,  while the Cauchy horizon is mapped to a horizon, as before.  This looks like the interior of an uncharged black hole. 

It would thus appear that at least in these gauges,  it is only meaningful to talk about the exterior region in the dual geometry.   It would be nice to explore this aspect in more detail.

\section{Classification of Higher-Derivative Corrections and Non-Perturbative Solutions}\label{sec-solun}

We have so far restricted the discussion to the two-derivative spacetime theory \eqref{act1}.  We would now like to explore the higher-derivative generalization of the effective one-dimensional action \eqref{action12} in the time-independent scenario \eqref{ansatz},  consistent with the duality properties and $x$-reparametrization invariance.  It is possible to write down the most general form of the higher-derivative action,
\baa
S^{(H)} = \frac{1}{16\pi G} \int \dd{x} n e^{-\Phi} F \pqty{ {\cal D} \Phi, {\cal D}^2 \Phi, \cdots; {\cal S} ,  {\cal D } {\cal S} , {\cal D}^2 {\cal S}, \cdots  }. \label{genhighdcorr}
\ea
This form of the action is unfortunately difficult to work with.  One can organize the terms by the number of derivatives and enumerate the terms.  In a given term, the derivatives are counted by the number of $\cal D$'s.   In \cite{Hohm:2019jgu}, the authors had shown that one can vastly simplify the form of the action by using appropriate field redefinitions.

The arguments presented in \S 2.2 of \cite{Hohm:2019jgu} (steps 1--5) go through \textit{mutatis mutandis} in our case as well.  
To be more specific, using the equations of motion for the fields $\Phi$ and $\cal S$ and using integration by parts (one has to be careful with the variation of ${\cal S}$ since it is a constrained field, ${\cal S}^2 =1$),  it is possible to show that the total action takes the form
\baa
S= \frac{1}{16\pi G} \int \dd{x} n e^{-\Phi} \pqty{ \ldc^2 +  (\cD \Phi)^2 + F (\cD \cS) }, \label{action121}
\ea 
where the function $F$ can be written down in a derivative expansion as,
\baa
F (\cD \cS)  &=  \frac18 \Tr (\cD \cS)^2 +  \pqty{ c_{4,1}  \Tr (\cD \cS)^4 + c_{4,2} \bqty{ \Tr (\cD \cS)^2 }^2 } \\
&\quad + \pqty{ c_{6,1}  \Tr (\cD \cS)^6 + c_{6,2}  \Tr (\cD \cS)^4 \Tr (\cD \cS)^2 + c_{6,3} \bqty{ \Tr (\cD \cS)^2 }^3     } \\
&\quad + \cdots. \label{highdcorr}
\ea
Here $c_{n,m}$'s  are Wilson coefficients associated with different terms appearing at the $n$\textsuperscript{th} order in derivative counting.  It is important in the argument that the leading order action is precisely the two-derivative action \eqref{action12}.  In obtaining the analogous form of the Lagrangian,  the authors of \cite{Hohm:2019jgu} had used only redefinitions of $\Phi$ and $\cS$. They were further able to simplify the higher-derivative terms using redefinitions of $n$ and eliminate all terms involving $\Tr (\cD \cS)^2$.  However,  as in \cite{Codina:2023fhy},  we want to preserve the measure $n e^{-\Phi}$ appearing in the action \eqref{action121} which multiplies $\ldc^2$.  Therefore,  the field redefinitions of $n$ are completely locked to that of $\Phi$ and we cannot perform any further redefinition.  However,  as we show now,  thanks to the fact that we are working in a low-dimensional scenario, this is immaterial in the present context and we can obtain an even stronger result.  It is obvious from \eqref{highdcorr} that the function can be characterized by the eigenvalues of the matrix $\cD \cS$.  By using the definitions \eqref{genmet1},  \eqref{defcalS},  \eqref{defcovD}, we find that the eigenvalues of $\cD \cS$ are given by $0,  \pm \mi \sqrt{2} \eig$ where
\baa
\eig^2 \equiv \frac{1}{n(x)^2 m(x)^2} \pqty{2m'(x)^2 - V'(x)^2  }.  \label{eigdef}
\ea
This immediately implies that $F(\cD S)$ is actually a function of a \emph{single} variable $\eig$ and that at any even order $n$ in the derivatives,  the Wilson coefficients $c_{n,m}$ are all related with a single term $\eig^n$.  For instance,  for the four-derivative term,  we would have $\Tr (\cD \cS)^4 = \frac12 \bqty{ \Tr (\cD \cS)^2  }^2 = 8 \eig^4$.  This allows us to write the function $F(\eig)$ in the form
\baa
F(\eig) = - \frac{1}{2} \eig^2 + \sum_{n=2}^\infty c_{2n} \eig^{2n}.  \label{feig}
\ea
It is quite remarkable that the field strength and the curvature are tied up in the expansion in a very coordinated way as in \eqref{eigdef},  with only one independent Wilson coefficient appearing at each order in derivative.  This is rather surprising from the point of view of low-energy effective field theory; more so when one considers $r$ gauge fields,  see \S\ref{sec-discussion}.

It might be tempting to interpret \eqref{feig} as an expansion in $\alpha'$, but as carefully elucidated in \cite{Codina:2023fhy} one must be cautious in the context of non-critical strings:  this derivative expansion is sensible only when the coefficients $c_{2n}$, suitably non-dimensionalized, are parametrically suppressed compared to the previous orders. This approach is well-suited for using perturbation theory,  but one can adopt the point of view \cite{Codina:2023nwz} that we have a fully \emph{non}-perturbative theory in $\alpha'$ with an arbitrary $F(\eig)$ satisfying the minimal conditions that 
\begin{enumerate}
\item  It is an even function of the eigenvalue,  $F(\eig) = F(-\eig)$.
\item  The leading term of the Taylor  expansion of the function about $\eig = 0$ is that of the two-derivative theory,  $F(\eig) = -\eig^2/2 + \mathcal{O} (\eig^4)$.
\end{enumerate}

\subsection{The Non-Perturbative Solution}\label{subsec-nonp}

We now work with the full action
\baa
S= \frac{1}{16\pi G} \int \dd[2]{x} n e^{-\Phi} \pqty{ \ldc^2 +  (\cD \Phi)^2 + F (\eig) }, \label{action122}
\ea 
with $\eig$ as defined by \eqref{eigdef} and $F(\eig)$ having the aforementioned properties. The equations of motion for $n$,  $\Phi$,  $m$, $V$ as defined previously are as follows:
\baa
E_n &= e^{-\Phi} \bqty{ \ldc^2 - (\cD \Phi)^2  + F(\eig) - \eig F'(\eig) } = 0, \\
E_\Phi &= - n e^{-\Phi} \bqty{ \ldc^2  -  (\cD \Phi)^2    + F(\eig)  + 2 \cD^2 \Phi} = 0, \\
E_m &= - \frac{n}{m} e^{-\Phi} \eig F'(\eig) -2 n\cD \pqty{  \frac{e^{-\Phi} }{m^2 \eig} F'(\eig) \cD m  } = 0,\\
E_V &= n \cD \pqty{  \frac{e^{-\Phi} }{m^2 \eig} F'(\eig) \cD V  }  = 0.  \label{eomgen}
\ea
The solution to this set of equations would appear to be hopelessly difficult unless some perturbative method is adopted.  However, using a parametrization introduced by \cite{Gasperini:2023tus}, the authors of \cite{Codina:2023nwz} were able to solve the equations exactly in the case involving no gauge fields. We will see that even in our case, we will be able to solve the equations exactly even though the form of the equations are more complicated than those in \cite{Codina:2023nwz}.  The authors of \cite{Codina:2023nwz} utilized some symmetries in the equations of motion to obtain the solution. Quite surprisingly,  even in the absence of such symmetries, we can obtain the solution in a very similar form.

Translated to the case at hand,  the parametrization of \cite{Gasperini:2023tus} proceeds from the simple but restrictive assumption that the function
\baa
f(\eig) \equiv F'(\eig) \label{gaspven}
\ea
admits a \emph{single-valued} inverse function $\eig(f)$.  We will be using $f$ as a coordinate to parametrize the solutions. Various important properties of the $f$-parametrization are described in detail in \cite{Codina:2023nwz}.

In this case,  it is convenient to work in the gauge $n(x) = 1$.  We will first work with the $E_n = 0$ equation in \eqref{eomgen}. We define
\baa
G(\eig) \equiv \eig F'(\eig) - F(\eig),   \label{geigdef}
\ea
from which it follows that $G(0) = 0 = G'(0)$. Using the previous definition \eqref{gaspven}, we can write the function in terms of the $f$ parameter,
\baa
G(f) = \int_0^f \dd{f'} \eig(f'). \label{gfintdef}
\ea
It then follows from the $E_n$ equation that
\baa
\Phi'(x)^2 = \ldc^2 - G(f).  \label{phix2eq1}
\ea
The combination $nE_n + E_\Phi$,  on the other hand,  yields,
\baa
2\Phi''(x) = - \eig(f) f \label{phi2xeq2}.
\ea
Taking the $x$-derivative of eq. \eqref{phix2eq1} and using both eqs. \eqref{phix2eq1} and \eqref{phi2xeq2} we obtain the expression for the coordinate $x$ in terms of $f$,
\baa
x =x_0  \pm \int^f \frac{\dd{f'}}{f' \sqrt{\ldc^2 - G(f') }},  \label{xingfgasven}
\ea
where the appropriate sign has to be chosen based on the physics it describes.  Using this relation between the coordinates $x$ and $f$,  we find that $\Phi'(f)^2 = f^{-2}$,  as a result of which we have
\baa
\Phi (f) = \log \frac{f}{\ldc} + \Phi_0. \label{phifsol1}
\ea
There was a possible sign ambiguity in front of the logarithm in the solution, which we fixed by using eq.  \eqref{phi2xeq2}.  Furthermore,  we could have taken the absolute value of $f$ within the logarithm,  but we can always restrict to a positive $f$, or more precisely $\Re f > 0$,   without any loss of generality.   To make the argument of the logarithm dimensionless,  we have also chosen to divide by $\ldc$,  an overall rescaling of which can be absorbed in the additive constant $\Phi_0$.

At this stage,  it is also worth emphasizing that we have been able to determine the functions $x(f)$ and $\Phi(f)$ by using only two of the four equations of motion \eqref{eomgen}.  The authors of \cite{Codina:2023nwz} could use a property of the $E_m$ equation to simplify the analysis --- while this property is not shared by the equation in our case,  we have shown that fewer equations actually suffice to determine the forms $x(f)$ and $\Phi(f)$. 

Furthermore,  rather surprisingly, the forms of these functions are the same as in the uncharged case \cite{Codina:2023nwz}.  It is surprising because, for instance, the duality-invariant dilaton $\Phi$, in the Schwarzschild-Droste--type $x$ coordinates has different functional forms explicitly depending on the charge and mass,  see \eqref{origsols}.  In this regard,  the behavior of $\Phi$ with respect to the $f$ coordinate is analogous to the dependence of the string dilaton $\phi$ on the coordinate $x$ in the gauge $n(x) = 1/m(x)$.  It is also made obvious by eq. \eqref{phifsol1} that it is actually the duality-invariant dilaton that is being used as a coordinate in the full non-perturbative $f$-parametrization.

We next use the gauge field equation $E_V = 0$, which yields
\baa
\frac{e^{-\Phi} f V'(x) }{\eig(f) m(x)^2 } = {\cal Q} \ldc e^{-\Phi_0}, \label{fullthechar}
\ea
where ${\cal Q}$ is a constant, related to the total charge, as before.
Using the solution \eqref{phifsol1} in this relation,  we find that
\baa
V'(x) = {\cal Q} \eig m^2.  \label{vprimefull}
\ea
At this point, we use the definition of $\eig$,  \eqref{eigdef} and the relation \eqref{xingfgasven} to write a differential equation for $m(f)$,
\baa
m'(f)^2 = \frac12 m(f)^2 (1 + {\cal Q}^2 m(f)^2)  \frac{\eig(f)^2}{f^2 \pqty{\ldc^2  - G(f) }},  \label{mfode}
\ea
which can be solved to yield
\baa
m(f)^2 = \frac{1}{{\cal Q}^2} \csch^2 \qty( \int_{f_m}^f \dd{f'} \frac{\eig(f')}{\sqrt{2} f' \sqrt{\ldc^2 - G(f')}}  ).  \label{mfullqsol}
\ea
We have introduced another constant of integration $f_m$ inside the $\csch$ function, which is related to the mass in this spacetime.  As expected,  the form of the solution \eqref{mfullqsol} is markedly different from the neutral counterpart \cite{Codina:2023nwz}.  In particular,  we cannot take the na\"ive uncharged limit $ {\cal Q} \to 0$,  to obtain the corresponding solution.

All that remains now is the solution to the gauge field equation,  which can be done easily after using \eqref{xingfgasven} and \eqref{mfullqsol},
\baa
V(f) = V_0 \pm {\cal Q} \int^f \dd{f'} \frac{\eig(f') m(f')^2}{f' \sqrt{\ldc^2 - G(f')}}.  \label{vfullqsol}
\ea

In summary,  \eqref{xingfgasven},  \eqref{phifsol1},  \eqref{mfullqsol}, \eqref{vfullqsol} and the gauge condition $n(x) = 1$ describe the full non-perturbative solution to the set of equations \eqref{eomgen}.

\subsection{Charged Black Holes in the Non-Perturbative Parametrization}\label{subsec-chargbhnp}

The authors of \cite{Codina:2023nwz} discussed many aspects of the non-perturbative parametrization of \cite{Gasperini:2023tus}.  We will now  explore some salient features of this parametrization in the context of charged black holes and also point out some new subtleties that arise.

Note from the previous subsection that to obtain the solution in terms of the $f$-parametrization,  we used the gauge $n(x) = 1$.  In this case,   the continuation of the coordinate across horizons is somewhat involved and the best one can do is to patch together solutions in different spacetime regions.  We can generalize the analytic continuation prescription of \cite{Codina:2023nwz} to describe the solution in the black hole interior: we make the replacements
\baa
m(x) \to \mi m(x), \quad n(x) \to -\mi n(x), \quad \Phi(x) \to \Phi(x) - \frac{\mi\pi}{2},  \quad V(x) \to \mi V(x) \label{replacean}
\ea
in the full action \eqref{action122}.  The transformation rule for the gauge field is determined by how $m$ transforms since these fields appear together in the generalized metric \eqref{genmet1}.  This  continuation entails a number of sign changes in the equation of motion, but they can be solved in a similar way as in \S\ref{subsec-nonp}.  We can generalize the procedure outlined in \cite{Codina:2023nwz}  is as follows: we consider different regions ${\cal R}_1$,  ${\cal R}_2$, $\cdots$, where $x$ alternates between these regions as a spacelike and timelike coordinate.  One must pay attention to the branch cut structure in the complex $f$-plane and 
we can  then solve the equations of motion in different regions,  but to match the solutions at the interface of the regions we must look at the \emph{finite} scalar invariants --- the string dilaton $\phi$,  the curvature scalar $R$ and the electric field strength $F_{\mu \nu} F^{\mu \nu}$ --- rather than the coordinates themselves.

One can ask whether it is possible to describe exactly extremal black holes in this formalism,   with degenerate horizons. In this case,  we can think about two regions: the black hole exterior (${\cal R}_1$) and the region between the singularity and the horizon (${\cal R}_2$).  We need not perform the analytic continuation \eqref{replacean}. However,  it is known from experience that such limits can often be rather singular,  let us see if that is the case here.  For definiteness,  let us consider the two-derivative theory.  It is useful to rewrite the solution \eqref{origsols} in the $n(x) = 1$ gauge.  It is straightforward to show that the (exterior) solution takes the form,
\baa
\dd{s}^2 &= - \pqty{ \frac{e^{2 \ldc x} (\mu^2 - Q^2) - 1 }{e^{2 \ldc x} (\mu^2 - Q^2) + 2\mu e^{ \ldc x}  + 1 } }^2 \dd{t}^2 + \dd{x}^2, \\
\Phi (x) &= -\log( e^{\ldc x} (\mu^2 - Q^2) - e^{-\ldc x} ) + \log 2, \\
V(x) &= - \frac{2\sqrt{2} Q}{e^{\ldc x} (\mu^2 - Q^2) + 2\mu + e^{ -\ldc x}  }. \label{cbhnx1}
\ea
We now find the solution in the $f$-parametrization. In the two-derivative theory, we have $F(\eig) = - \eig^2/2$ and thus it follows from the definitions \eqref{gaspven} and \eqref{gfintdef} that
\baa
\eig(f) = -f,  \quad G(f) = - \frac12 f^2.  \label{eigfgf}
\ea
Using this form of $G(f)$ in eq.  \eqref{xingfgasven} yields, with the lower sign,
\baa
x = x_0 + \frac{1}{\ldc} \csch^{-1} \frac{f}{\sqrt{2} \ldc},  \label{xf2dersol}
\ea
which leads to the dilaton solution \eqref{phifsol1}
\baa
\Phi = \Phi_0  - \log (  \sqrt{2} \sinh( \ldc (x-x_0) )  )  \label{phif2derfsol}.
\ea
Quite happily,  the functional forms of $\Phi$ in \eqref{cbhnx1} and \eqref{phif2derfsol} match up and we can find the identification,
\baa
x_0 &= - \frac{1}{2\ldc} \log(\mu^2 - Q^2), \\
\Phi_0 &=  - \frac{1}{2} \log \frac{\mu^2 - Q^2}{2}. \label{matchcohns}
\ea
Not as happily,  we note that in the extremal limit $\mu \to Q$,  the constants $x_0$ and $\Phi_0$ are divergent quantities.  The concern that the extremal limit could be singular turns out to be realized for this parametrization. The  remedy,  however, is simple: we should \emph{not} consider the exactly extremal scenario to begin with,  but the slightly non-extremal one to tame the possible singularities. This also necessitates the inclusion of the region between the Cauchy and event horizons.  We can then describe the whole geometry using the analytic continuation \eqref{replacean} in this region.

We can write down the solution for $m^2$ in the $f$-parametrization using \eqref{mfullqsol},
\baa
m(f)^2 = \frac{4\ldc^2}{{\cal Q}^2 \pqty{f \sqrt{f_m^2 + 2 \ldc^2} - f_m  \sqrt{f^2 + 2 \ldc^2} }^2 },  \label{mf2derfull}
\ea
matching which with the known solution \eqref{cbhnx1} yields the parameter identifications,
\baa
\mu &= - \frac{\sqrt{f_m^2 + 2 \ldc^2 }}{f_m} e^{-\ldc x_0},  \\  {\cal Q}^2 &=  \frac{2\ldc^2}{f_m^2} = \frac{Q^2}{ \mu^2 - Q^2}. \label{furtheridentif1}
\ea
To complete the discussion, we also write down the solution for the gauge field in this parametrization,  with a constant subtracted to make it vanish at $f=0$ (asymptotic infinity),
\baa
V(f) = \frac{\sqrt{2} f  \pqty{f \sqrt{(f^2+2 \ldc ^2) (f_m^2+2 \ldc ^2)}+f^2 f_m+2 \ldc^2 f_m} }{{\cal Q} f_m (f^2-f_m^2) \sqrt{f^2+2 \ldc ^2}}.  \label{vf2derfull}
\ea
This matches exactly with the previous solution with the identifications that have been made already.

Let us comment on another novel feature of this set-up: the region between the singularity and the Cauchy horizon.  In this region,  the coordinate $x$ has the same nature as in the black hole exterior.  Therefore,  it might appear that in the $f$-parametrization,  the same solutions \eqref{phifsol1},  \eqref{mf2derfull} and \eqref{vf2derfull} would apply in this region as well.  Given the exterior solution is regular,  where then is the singularity?  The answer lies with the constants of integration. In the general procedure using the $f$-parametrization, we have, to begin with,  independent constants in each region and we can fix these constants during the matching procedure at the interfaces.  From the form of the solution of the metric component \eqref{mfullqsol},  we see that the metric diverges at $f = f_m$,  (ignoring for a moment the other possible zeros of the integral inside the $\csch^2$ function),  leading to a curvature singularity (we have to use eq.  \eqref{xingfgasven} as well).  We see that the metric component \eqref{mf2derfull} is indeed singular around $f = f_m$. How,  then,  is this a regular solution? Recall that we have posited that a single patch is described only by the region $\Re f > 0$.  Since for a positive mass spacetime $f_m < 0$ from \eqref{furtheridentif1},  this region does not have the singularity. 

Indeed,  we see that in this case,  the metric component is given by
\baa
m(x)^2 = \pqty{ \frac{1- (\mu^2 -Q^2) e^{2\ldc x} }{ 2 \mu e^{\ldc x} - 1 - (\mu^2 -Q^2) e^{2\ldc x}  } }^2,  \label{mxbehindch}
\ea
and in the $f$-parametrization,  we have to choose the lower sign in \eqref{xingfgasven}, eventually leading to the same constants as \eqref{matchcohns} in this region. However,  now the matching procedure yields
\baa
\mu &=  \frac{\sqrt{f_m^2 + 2 \ldc^2 }}{f_m} e^{-\ldc x_0},    \label{mubehindch}
\ea
with $f_m > 0$.  Thus,  the singular point $f = f_m$ lies on the right half-plane.  
Note that, unlike in the neutral case,  the singularity is at a finite point in the $f$-parametrization.
This could also have been anticipated based on our discussion of duality in \S\ref{subsec-tdual}. 

It would be very interesting to explore other interesting aspects of this parametrization for charged black holes in future works.

\section{Discussion}\label{sec-discussion}

It is quite satisfying that the classification program of \cite{Hohm:2019jgu} for higher-derivative corrections can be generalized to the  heterotic effective theory in two dimensions with an $\mathrm{O}(1, 2; \mathbb{R})$ symmetry.  Quite surprisingly,  as shown in \S\ref{subsec-nonp}, the non-perturbative parametrization of the classical solution \cite{Gasperini:2023tus, Codina:2023nwz} can be extended to the heterotic case as well.  Although there are two fields (metric and gauge field) involved in the higher-derivative expansion,  the solution could be parametrized in terms of a single parameter since the derivative $\cD \cS$ has just one independent eigenvalue. This motivates us to wonder whether there is something special about just one abelian gauge field or whether this structure persists more generally.  

Let us then consider the most general two-dimensional abelian case with $r$ gauge fields (we do not have the freedom to add more metric components or the Kalb-Ramond field and the non-abelian gauge fields do not exhibit the symmetry).  In other words, we consider $r$ abelian factors in the Cartan of the heterotic gauge group,  which means that $r \leq 12$,  but the specific value of $r$ will not matter in the following discussion.  In this case,  the symmetry group of classical string theory is $\mathrm{O}(1, 1 +r; \mathbb{R})$.  We can assume a similar time-independent structure as in \eqref{ansatz},  now with the gauge fields $A_{\mu}^{ (a)} \dd{x^\mu} = V^a (x) \dd{t}$, where $a (= 1, 2,  \cdots, r)$ labels the Cartan components,   which has the following structure of the generalized metric \cite{Hohm:2011ex},  written in a $(2,r)$ block form,
\baa
{\cal H} =\pqty{ \begin{array}{cc:c}
- \dfrac{1}{m(x)^2} & \quad \dfrac{\delta_{a b} V^a (x) V^b (x) }{2 m(x)^2}  & \quad \dfrac{V^a (x) }{m(x)^2}  \\[-2pt]
\dfrac{\delta_{a b} V^a (x) V^b (x) }{2 m(x)^2}  & \quad -\dfrac{\pqty{2m(x)^2 - \delta_{a b} V^a (x) V^b (x)}^2}{4m(x)^2} & \quad V^a (x) \pqty{1 - \dfrac{\delta_{b c} V^{b}  V^c  }{2m(x)^2} }  \\[10pt]
\hdashline\\[-10pt]
\dfrac{V^a (x) }{m(x)^2}  & \quad V^a (x) \pqty{1 - \dfrac{\delta_{b c} V^{b} (x) V^c (x) }{2m(x)^2} }  &\quad \delta^{a b} -\dfrac{V^a(x) V^b (x)}{m(x)^2}
\end{array} }, \label{genmetgen2r}
\ea
where the Kronecker delta $\delta_{a b}$ is  the Killing-Cartan metric in the space of gauge fields. The $\mathrm{O}(1, 1 +r; \mathbb{R})$-invariant metric is now given by $\eta = \sigma_1 \oplus \bm{1}_r$, where $\sigma_1$ is the first Pauli matrix. Defining $\cS$ as in \eqref{defcalS},  we find that ${\cD \cS}$ is a rank-2 $(2+r)\times (2+r)$ matrix with $r$ zero eigenvalues and 2 non-zero eigenvalues $\pm\mi \sqrt{2} \eig$ with
\baa
\eig^2= \frac{1}{n(x)^2 m(x)^2} \pqty{2m'(x)^2 - \delta^{a b} V^{'}_{a}(x) V^{'}_{b}(x)}.  \label{geneig2r}
\ea
This is exactly the structure we needed for the $f$-parametrization.  Therefore,  not only would the higher-derivative corrections have the same form as  \eqref{feig},  the non-perturbative solution of \S\ref{subsec-nonp} would go through in this more general case as well.

There are several other interesting aspects which we can explore in the heterotic context.  Just as the two-derivative uncharged black hole is related to an $ {\mathrm{SL} (2, \mathbb{R})}/{\mathrm{U} (1)}$ gauged Wess-Zumino-Witten (WZW) model \cite{Witten:1991yr},  the heterotic black hole has a corresponding relation to an $ \pqty{\mathrm{SL} (2, \mathbb{R}) \times \mathrm{U} (1)}/{\mathrm{U} (1)}$ gauged WZW model (see \cite{Giveon:2003ge}, for example).  It would be very interesting to examine whether powerful tools of representation theory can be used for gaining better insights into the aspects of two-dimensional black holes investigated in this article, including higher-derivative corrections and non-perturbative solutions.  An extension of the results described in \cite{Lunin:2024vsx} in the present context would also be instructive.

A particularly interesting physics question concerns singularities.  Working with the generalized metric formalism inspired from double field theory,  we exhibited the precise map of the metric and gauge field between different duality frames.  As explained in \S\ref{subsec-gauge},  an interesting issue arises vis-\`{a}-vis the gauge choice and duality.  While we posited a particular gauge fixing which is physically reasonable,  it would be more satisfactory to gain a better understanding of this issue from a more fundamental point of view.  We pointed out some very interesting features of the singularity in the original and dual geometries.  We noted that duality gives a geometry with the same charge, but opposite mass.  Furthermore,  if we start out from asymptotic infinity,  we generally encounter a timelike singularity,  even though it is an infinite affine distance away.  

While we considered time-independent configurations,  it would be interesting to frame dynamical questions in these backgrounds.  Since the original geometry has Cauchy horizons,  one could wonder whether the strong cosmic censorship conjecture, which implies an instability of the Cauchy horizon,  would hold true.  The results in \cite{Moitra:2020ojo,  MoitraThesis},  supplemented with a proper matching in the asymptotically flat exterior would seem to imply that the conjecture would hold true for the black holes discussed in this paper even in the near-extremal case.  More troublingly,    the dual geometry seems to sit uneasily with the weak cosmic censorship conjecture which forbids naked singularities.  It would be worthwhile to explore such dynamical questions in a thorough manner, consistent with the constraints of string theory.

As mentioned previously,  the $\mathrm{O}(1, 2;  \mathbb{R})$ symmetry for a single gauge field is expected to be broken to $\mathrm{O}(1, 2;  \mathbb{Z})$ by quantum effects.  The group $\mathrm{PSL} (2, \mathbb{Z})$, which sits naturally within $\mathrm{O}(1, 2 ;  \mathbb{Z})$ as a subgroup,  has been known to be important in string theory in many ways. It would also be interesting to consider the full non-abelian heterotic theory,  which shows a reduced symmetry $\mathrm{O}(1, 1;  \mathbb{R})$ at the classical level.   There are other intriguing possibilities to consider in this broader framework, such as the question of quantum entanglement in heterotic  string backgrounds \cite{Dabholkar:2024neq},  which is known to exhibit some unique features.   We hope to address such exciting questions in future works.

\acknowledgments

I would like to acknowledge the hospitality of the Simons Center for Geometry and Physics, Stony Brook University at the 2025 Simons Physics Summer Workshop,  during an early stage of this work. 
This work was performed in part at the Aspen Center for Physics, which is supported by a grant from the Simons Foundation (1161654, Troyer).
Various stages of this research were supported by the European Research Council (ERC) under the European Union's  Seventh Framework Programme (FP7/2007-2013), ERC Grant Agreement ADG 834878,   and by the European Union's  Horizon 2020 Research and Innovation Programme (Grant Agreement No. 101115511).

 \bibliographystyle{JHEP}
 \bibliography{hetbh_refs}

\providecommand{\href}[2]{#2}\begingroup\raggedright\begin{thebibliography}{10}

\bibitem{Giveon:1994fu}
A.~Giveon, M.~Porrati and E.~Rabinovici, \emph{{Target space duality in string
  theory}}, \href{https://doi.org/10.1016/0370-1573(94)90070-1}{\emph{Phys.
  Rept.} {\bfseries 244} (1994) 77}
  [\href{https://arxiv.org/abs/hep-th/9401139}{{\ttfamily hep-th/9401139}}].

\bibitem{Buscher:1987sk}
T.H.~Buscher, \emph{{A Symmetry of the String Background Field Equations}},
  \href{https://doi.org/10.1016/0370-2693(87)90769-6}{\emph{Phys. Lett. B}
  {\bfseries 194} (1987) 59}.

\bibitem{Buscher:1987qj}
T.H.~Buscher, \emph{{Path Integral Derivation of Quantum Duality in Nonlinear
  Sigma Models}},
  \href{https://doi.org/10.1016/0370-2693(88)90602-8}{\emph{Phys. Lett. B}
  {\bfseries 201} (1988) 466}.

\bibitem{Brandenberger:1988aj}
R.H.~Brandenberger and C.~Vafa, \emph{{Superstrings in the Early Universe}},
  \href{https://doi.org/10.1016/0550-3213(89)90037-0}{\emph{Nucl. Phys. B}
  {\bfseries 316} (1989) 391}.

\bibitem{Tseytlin:1991xk}
A.A.~Tseytlin and C.~Vafa, \emph{{Elements of string cosmology}},
  \href{https://doi.org/10.1016/0550-3213(92)90327-8}{\emph{Nucl. Phys. B}
  {\bfseries 372} (1992) 443}
  [\href{https://arxiv.org/abs/hep-th/9109048}{{\ttfamily hep-th/9109048}}].

\bibitem{Veneziano:1991ek}
G.~Veneziano, \emph{{Scale factor duality for classical and quantum strings}},
  \href{https://doi.org/10.1016/0370-2693(91)90055-U}{\emph{Phys. Lett. B}
  {\bfseries 265} (1991) 287}.

\bibitem{Meissner:1991zj}
K.A.~Meissner and G.~Veneziano, \emph{{Symmetries of cosmological superstring
  vacua}}, \href{https://doi.org/10.1016/0370-2693(91)90520-Z}{\emph{Phys.
  Lett. B} {\bfseries 267} (1991) 33}.

\bibitem{Sen:1991zi}
A.~Sen, \emph{{O(d) x O(d) symmetry of the space of cosmological solutions in
  string theory, scale factor duality and two-dimensional black holes}},
  \href{https://doi.org/10.1016/0370-2693(91)90090-D}{\emph{Phys. Lett. B}
  {\bfseries 271} (1991) 295}.

\bibitem{Gasperini:1991ak}
M.~Gasperini and G.~Veneziano, \emph{{O(d,d) covariant string cosmology}},
  \href{https://doi.org/10.1016/0370-2693(92)90744-O}{\emph{Phys. Lett. B}
  {\bfseries 277} (1992) 256}
  [\href{https://arxiv.org/abs/hep-th/9112044}{{\ttfamily hep-th/9112044}}].

\bibitem{Maharana:1992my}
J.~Maharana and J.H.~Schwarz, \emph{{Noncompact symmetries in string theory}},
  \href{https://doi.org/10.1016/0550-3213(93)90387-5}{\emph{Nucl. Phys. B}
  {\bfseries 390} (1993) 3}
  [\href{https://arxiv.org/abs/hep-th/9207016}{{\ttfamily hep-th/9207016}}].

\bibitem{Siegel:1993th}
W.~Siegel, \emph{{Superspace duality in low-energy superstrings}},
  \href{https://doi.org/10.1103/PhysRevD.48.2826}{\emph{Phys. Rev. D}
  {\bfseries 48} (1993) 2826}
  [\href{https://arxiv.org/abs/hep-th/9305073}{{\ttfamily hep-th/9305073}}].

\bibitem{Hull:2009mi}
C.~Hull and B.~Zwiebach, \emph{{Double Field Theory}},
  \href{https://doi.org/10.1088/1126-6708/2009/09/099}{\emph{JHEP} {\bfseries
  09} (2009) 099} [\href{https://arxiv.org/abs/0904.4664}{{\ttfamily
  0904.4664}}].

\bibitem{Aldazabal:2013sca}
G.~Aldazabal, D.~Marques and C.~Nunez, \emph{{Double Field Theory: A
  Pedagogical Review}},
  \href{https://doi.org/10.1088/0264-9381/30/16/163001}{\emph{Class. Quant.
  Grav.} {\bfseries 30} (2013) 163001}
  [\href{https://arxiv.org/abs/1305.1907}{{\ttfamily 1305.1907}}].

\bibitem{Hohm:2019jgu}
O.~Hohm and B.~Zwiebach, \emph{{Duality invariant cosmology to all orders in
  $\alpha$'}}, \href{https://doi.org/10.1103/PhysRevD.100.126011}{\emph{Phys.
  Rev. D} {\bfseries 100} (2019) 126011}
  [\href{https://arxiv.org/abs/1905.06963}{{\ttfamily 1905.06963}}].

\bibitem{Codina:2023fhy}
T.~Codina, O.~Hohm and B.~Zwiebach, \emph{{2D black holes, Bianchi I
  cosmologies, and {\ensuremath{\alpha'}}}},
  \href{https://doi.org/10.1103/PhysRevD.108.026014}{\emph{Phys. Rev. D}
  {\bfseries 108} (2023) 026014}
  [\href{https://arxiv.org/abs/2304.06763}{{\ttfamily 2304.06763}}].

\bibitem{Elitzur:1990ubs}
S.~Elitzur, A.~Forge and E.~Rabinovici, \emph{{Some global aspects of string
  compactifications}},
  \href{https://doi.org/10.1016/0550-3213(91)90073-7}{\emph{Nucl. Phys. B}
  {\bfseries 359} (1991) 581}.

\bibitem{Rocek:1991vk}
M.~Ro\v{c}ek, K.~Schoutens and A.~Sevrin, \emph{{Off-shell WZW models in
  extended superspace}},
  \href{https://doi.org/10.1016/0370-2693(91)90057-W}{\emph{Phys. Lett. B}
  {\bfseries 265} (1991) 303}.

\bibitem{Mandal:1991tz}
G.~Mandal, A.M.~Sengupta and S.R.~Wadia, \emph{{Classical solutions of
  two-dimensional string theory}},
  \href{https://doi.org/10.1142/S0217732391001822}{\emph{Mod. Phys. Lett. A}
  {\bfseries 6} (1991) 1685}.

\bibitem{Witten:1991yr}
E.~Witten, \emph{{On string theory and black holes}},
  \href{https://doi.org/10.1103/PhysRevD.44.314}{\emph{Phys. Rev. D} {\bfseries
  44} (1991) 314}.

\bibitem{Giveon:1991sy}
A.~Giveon, \emph{{Target space duality and stringy black holes}},
  \href{https://doi.org/10.1142/S0217732391003316}{\emph{Mod. Phys. Lett. A}
  {\bfseries 6} (1991) 2843}.

\bibitem{Tseytlin:1991wr}
A.A.~Tseytlin, \emph{{Duality and dilaton}},
  \href{https://doi.org/10.1142/S021773239100186X}{\emph{Mod. Phys. Lett. A}
  {\bfseries 6} (1991) 1721}.

\bibitem{Dijkgraaf:1991ba}
R.~Dijkgraaf, H.L.~Verlinde and E.P.~Verlinde, \emph{{String propagation in a
  black hole geometry}},
  \href{https://doi.org/10.1016/0550-3213(92)90237-6}{\emph{Nucl. Phys. B}
  {\bfseries 371} (1992) 269}.

\bibitem{Kiritsis:1991zt}
E.B.~Kiritsis, \emph{{Duality in gauged WZW models}},
  \href{https://doi.org/10.1142/S0217732391003341}{\emph{Mod. Phys. Lett. A}
  {\bfseries 6} (1991) 2871}.

\bibitem{McGuigan:1991qp}
M.D.~McGuigan, C.R.~Nappi and S.A.~Yost, \emph{{Charged black holes in
  two-dimensional string theory}},
  \href{https://doi.org/10.1016/0550-3213(92)90039-E}{\emph{Nucl. Phys. B}
  {\bfseries 375} (1992) 421}
  [\href{https://arxiv.org/abs/hep-th/9111038}{{\ttfamily hep-th/9111038}}].

\bibitem{Hohm:2011ex}
O.~Hohm and S.K.~Kwak, \emph{{Double Field Theory Formulation of Heterotic
  Strings}}, \href{https://doi.org/10.1007/JHEP06(2011)096}{\emph{JHEP}
  {\bfseries 06} (2011) 096} [\href{https://arxiv.org/abs/1103.2136}{{\ttfamily
  1103.2136}}].

\bibitem{Codina:2023nwz}
T.~Codina, O.~Hohm and B.~Zwiebach, \emph{{Black hole singularity resolution in
  D=2 via duality-invariant {\ensuremath{\alpha'}} corrections}},
  \href{https://doi.org/10.1103/PhysRevD.108.126006}{\emph{Phys. Rev. D}
  {\bfseries 108} (2023) 126006}
  [\href{https://arxiv.org/abs/2308.09743}{{\ttfamily 2308.09743}}].

\bibitem{Gasperini:2023tus}
M.~Gasperini and G.~Veneziano, \emph{{Non-singular pre-big bang scenarios from
  all-order {\ensuremath{\alpha}}' corrections}},
  \href{https://doi.org/10.1007/JHEP07(2023)144}{\emph{JHEP} {\bfseries 07}
  (2023) 144} [\href{https://arxiv.org/abs/2305.00222}{{\ttfamily
  2305.00222}}].

\bibitem{Moitra:2022umq}
U.~Moitra, \emph{{Self-similar gravitational dynamics, singularities and
  criticality in 2D}},
  \href{https://doi.org/10.1007/JHEP06(2023)194}{\emph{JHEP} {\bfseries 06}
  (2023) 194} [\href{https://arxiv.org/abs/2211.01394}{{\ttfamily
  2211.01394}}].

\bibitem{Giveon:1991jj}
A.~Giveon and M.~Ro\v{c}ek, \emph{{Generalized duality in curved string
  backgrounds}},
  \href{https://doi.org/10.1016/0550-3213(92)90518-G}{\emph{Nucl. Phys. B}
  {\bfseries 380} (1992) 128}
  [\href{https://arxiv.org/abs/hep-th/9112070}{{\ttfamily hep-th/9112070}}].

\bibitem{Giveon:2003ge}
A.~Giveon, E.~Rabinovici and A.~Sever, \emph{{Beyond the singularity of the 2-D
  charged black hole}},
  \href{https://doi.org/10.1088/1126-6708/2003/07/055}{\emph{JHEP} {\bfseries
  07} (2003) 055} [\href{https://arxiv.org/abs/hep-th/0305140}{{\ttfamily
  hep-th/0305140}}].

\bibitem{Lunin:2024vsx}
O.~Lunin and P.~Shah, \emph{{Double Field Theory and {\ensuremath{\alpha}}'
  corrections: Explicit examples}},
  \href{https://doi.org/10.1016/j.nuclphysb.2025.116932}{\emph{Nucl. Phys. B}
  {\bfseries 1017} (2025) 116932}
  [\href{https://arxiv.org/abs/2408.04833}{{\ttfamily 2408.04833}}].

\bibitem{Moitra:2020ojo}
U.~Moitra, \emph{{Strong Cosmic Censorship in Two Dimensions}},
  \href{https://doi.org/10.1103/PhysRevD.103.L081502}{\emph{Phys. Rev. D}
  {\bfseries 103} (2021) L081502}
  [\href{https://arxiv.org/abs/2011.03499}{{\ttfamily 2011.03499}}].

\bibitem{MoitraThesis}
U.~Moitra, \emph{{Aspects of Quantum Gravity, Holography and Entanglement}},
  Ph.D. thesis, Tata Institute of Fundamental Research.

\bibitem{Dabholkar:2024neq}
A.~Dabholkar and U.~Moitra, \emph{{Heterotic strings and quantum
  entanglement}}, \href{https://doi.org/10.1007/JHEP12(2024)012}{\emph{JHEP}
  {\bfseries 12} (2024) 012}
  [\href{https://arxiv.org/abs/2407.17553}{{\ttfamily 2407.17553}}].

\end{thebibliography}\endgroup

\end{document}